\documentclass{pnastwo}
\usepackage{graphicx}
\usepackage{amssymb}
\usepackage{cite}
\usepackage{amsmath}
\usepackage{enumitem}
\usepackage{caption}
\usepackage{bm}
\usepackage{pdfpages}
\usepackage{subcaption}
\DeclareMathOperator{\Var}{Var}
\begin{document}

\title{Harvesting entropy and quantifying the transition from noise to chaos in a photon-counting feedback loop}

\author{Aaron M. Hagerstrom\affil{1}{University of Maryland, College Park, Maryland, United States of America},
Thomas E. Murphy\affil{1}{},
\and
Rajarshi Roy\affil{1}{}}

\contributor{Submitted to Proceedings of the National Academy of Sciences of the United States of America}

\significancetext{The unpredictability of physical systems can depend on the scale at which they are observed. 
For example, single photons incident on a detector arrive at random times, but slow intensity variations can be observed 
by counting many photons over large time windows. We describe an experiment in which we modulate a weak optical 
signal using feedback from a single-photon detector. We quantitatively demonstrate a transition from single-photon shot noise to 
deterministic chaos. Furthermore, we show that measurements of the entropy rate of a system with small scale 
noise and large scale deterministic fluctuations can resolve both behaviors. We describe how quantifying entropy production can be used to evaluate physical random number generators.}

\maketitle

\begin{article}
\begin{abstract}
{Many physical processes, including the intensity fluctuations of a chaotic laser, the detection of single photons, and the 
Brownian motion of a microscopic particle in a fluid are unpredictable, at least on long timescales. This unpredictability can be due to a variety of 
physical mechanisms, but it is quantified by an entropy rate. This rate, which describes how quickly a system produces new and random information, 
is fundamentally important in statistical mechanics  and practically important for random number generation. We experimentally study entropy generation and the 
emergence of deterministic chaotic dynamics from discrete noise in a system that applies feedback to a weak optical signal at the single-photon level.
We show that the dynamics transition from shot noise to chaos as the photon rate increases, and that the entropy rate can reflect either the deterministic or noisy aspects
of the system depending on the sampling rate and resolution.}
\end{abstract}

\keywords{entropy | chaos | nonlinear dynamics | photon counting | statistical physics}

\dropcap{C}ontinuous variables and dynamical equations are often used to model systems whose time evolution is comprised of discrete events occurring at random times.
Examples include the flow of ions across cell membranes\cite{bialek2012biophysics}, the dynamics of large populations of neurons\cite{izhikevich2008large}, the birth and death of individuals in a population\cite{may2001}, traffic flow on roads\cite{chowdhury2000statistical}, 
the trading of securities in financial markets\cite{mantegna1999,preis2011switching}, 
infection and transmission of disease\cite{anderson1991}, and the emission and detection of photons\cite{mandel1995}.
We can identify two sources of unpredictability in these systems: the noise associated with the underlying random occurrences that comprise these signals, which is often 
described by a Poisson process, and the macroscopic dynamics of the system, which may be chaotic. When both effects are present, the macroscopic dynamics can alter the statistics of the noise, and the small-scale noise can in turn feed the large-scale dynamics.
This can lead to subtle and non-trivial effects including stochastic resonance and coherence resonance \cite{benzi1981mechanism,gammaitoni1998stochastic, pikovsky1997coherence}. Dynamical unpredictability and complexity are quantified by Lyapunov exponents and dimensionality, while shot noise is characterized by statistical metrics like average rate, variance, and signal-to-noise ratio. 
Characterizing the unpredictability of a system with both large-scale dynamics and small-scale shot noise remains an important challenge in many disciplines including statistical mechanics and information security.

Many cryptographic applications, including public key encryption\cite{lenstra2012} employ random numbers. 
Because the unpredictability of these numbers is essential, physical processes are sometimes 
used as a source of random numbers\cite{hamburg2012analysis, uchida2008,yamazaki2013,sakuraba2015,li2011, williams2010, rosin2013, kanter2009, harayama2012, sunada2012, oliver2011, oliver2013, virte2014,gabriel2010}. 
Physical random number generators are usually tested using the NIST\cite{rukhinnist2001} and Diehard\cite{marsaglia1998diehard}
test suites which assess their ability to produce bits that are free of bias and correlation. These tests 
are an excellent assessment of the performance of a physical random number generator in practical situations, 
but leave an important and fundamental problem unaddressed. Deterministic post-processing procedures, such as hash functions \cite{gabriel2010} are 
often employed to to remove bias and correlation. Because these procedures are algorithmic and reproducible, they cannot in principle increase the entropy rate of a bit stream. 
Thus, the reliability of a physical random number generator depends on an accurate assessment of the entropy rate of physical process that generated the numbers\cite{barker2012}. It
remains difficult to assess the unpredictability of a system based on physical principles. 

Evaluation of entropy rates from an information-theoretic perspective is also centrally important in statistical mechanics \cite{berut2012experimental,sethna2006statistical,jarzynski2015,touchette2009,roldan2010,andrieux2007,andrieux2008,caputo1987}.
One might expect that the unpredictability of a system with both small-scale shot noise and large-scale chaotic dynamics would depend on the scale at which it is observed.
In many systems, the dependence of the entropy rate on the resolution, $\varepsilon$, and the sampling interval, $\tau$, can reflect the 
physical origin of unpredictability \cite{gaspard1993,dettmann1999,cencini2000,boffetta2002}. This dependence has been been studied experimentally 
in Brownian motion, RC circuits, and Rayleigh-B\'{e}nard convection\cite{gaspard1993,gaspard1998,briggs2001tracking,andrieux2007,andrieux2008}. 

Here, we present an experimental exploration and numerical model of entropy production in a photon-counting optoelectronic 
feedback oscillator. Optoelectronic feedback loops which employ analog detectors and 
macroscopic optical signals produce rich dynamics whose timescales and dimensionality are highly tunable\cite{peil2009,murphy2010,larger2013,williams2013,chembo2005}. 
Our system applies optoelectronic feedback to a weak optical signal which is measured by a photon-counting detector.
The dynamic range of this system (eight orders of magnitude in time scale and a factor of 256 in photon rate) allows us 
to directly observe the transition from shot-noise dominated behavior to a low-dimensional chaotic attractor with increasing optical power -- a transition which 
to our knowledge has never been observed experimentally. We show that the entropy rate can reflect either the deterministic or stochastic aspects of the system,
depending on the sampling rate and measurement resolution and describe the importance of this observation for physical random number generation. 

\section{Experiment and Results}
\begin{figure}[htp]
 \centering
 \includegraphics{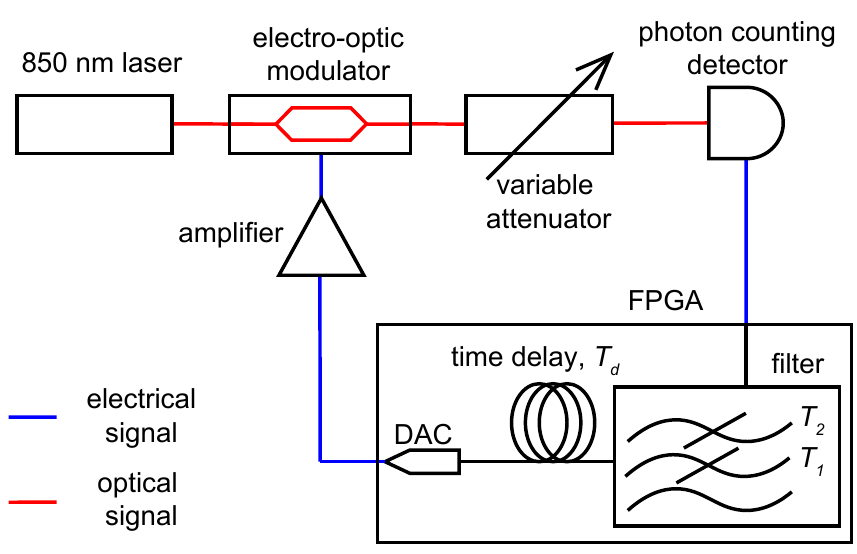}
 \caption{Experimental configuration. We employ a silicon avalanche photodiode, which detects individual photon arrivals. This signal is time-delayed and filtered using an FPGA, 
  and the output of the filter drives the modulator which in turn varies the light incident on the detector, forming a feedback loop.}\label{fig1}
\end{figure}

Figure 1 shows a schematic of our experimental configuration. Our system has a similar architecture to earlier experiments involving optoelectronic feedback loops,
but differs in that we use a photon-counting detector, whereas previous experiments used an analog photodetector. In either case, the signal from the detector is time-delayed and filtered, 
and the output of the filter drives the Mach-Zehnder electro-optic modulator (MZM) which in turn controls the light incident on the detector, 
forming a feedback loop. When an analog photodiode is used, the feedback loop is modeled by a time-delayed nonlinear differential equation.
\begin{equation}
\begin{aligned}
& \frac{d\bm{x}}{dt} = \bm{E}\bm{x} + \beta \bm{F} I(t) \\
& I(t)=\sin^2\left[ \bm{G}^T \bm{x}(t-T_d)+\phi \right] 
\end{aligned}\label{detEOM}
\end{equation}
Here, $\bm{x}$ is the state variable of a linear, time invariant filter, matrix $\bm{E}$ and the vectors $\bm{F}$ and $\bm{G}$ describe the characteristics
of the filter, $I(t)$ is the normalized intensity of light transmitted through the MZM, and $T_d$ is the time delay. When a photon-counting detector is used in place of an analog
photodiode, the filter variables can be modeled by a linear differential equation driven by discrete photon arrivals.
In our implementation, the equations of motion for the filter variables are
\begin{equation}
\frac{d}{dt}\begin{pmatrix} x_1 \\ x_2 \end{pmatrix} =  \begin{pmatrix}-\frac{1}{T_1} & 0 \\ 0 & -\frac{1}{T_2} \end{pmatrix}  \begin{pmatrix} x_1 \\ x_2 \end{pmatrix} +  \beta  \begin{pmatrix} 1 \\ 1 \end{pmatrix}  \frac{1}{\lambda_0} \sum_{i=1}^{\infty} \delta(t-t_i)   \\\label{randEOM},
\end{equation}
where the  photon arrivals times, $\{t_i\}$, are generated by a non-stationary Poisson point process whose rate, $\lambda(t)$, depends on the state of the filter variables. 
\begin{equation}
\lambda(t)=\lambda_0 I(t) = \lambda_0 \sin^2\left[ x_{1}(t-T_d)-x_{2}(t-T_d)+\phi \right]. \label{rate}
\end{equation}
In the limit that the $\lambda_0$ is large, the stochastic term in equation (\ref{randEOM}) can be replaced with its expectation value, $I(t)$, leading to
equation (\ref{detEOM}).

In our implementation, the time delay is $T_d$=1.734 ms, the modulator bias is $\phi=\pi/4$, and the filter constants $T_1$=1.2 ms, and $T_2$=60 $\mu$s. 
The filter and time delay are implemented digitally using an Altera Cyclone II field programmable gate array (FPGA) 
and a digital to analog converter (DAC). The clock speed of this device is 151.1515 MHz, and we record all of the photon arrival times to this 
precision. The light source in our experiment is a continuous wave fiber-coupled distributed feedback laser with a wavelength of 850 nm.
Our detector has a dark count rate of $\sim$ 100 counts/s and a dead time of about 40 ns. We vary the photon rate over a factor of 256, from $\lambda_0 T_d = 12.5$ ($7.20 \times 10^3$ count/s)
to $\lambda_0 T_d = 3200$ ($1.845\times 10^6$ counts/s). In all of the experiments shown here, $\beta$ is kept constant so that $\beta T_d = 8.87$.
\begin{figure}%[htp]
 %\centering
 \includegraphics{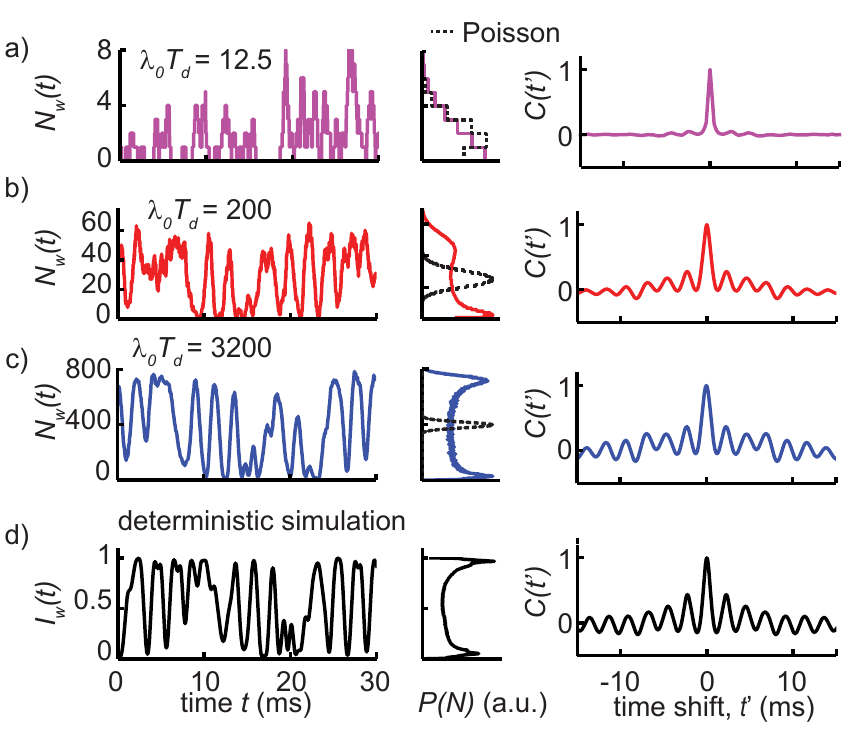}
 \caption{Time series, probability distributions, and autocorrelation functions. a) - c) show experimental data, and d) shows the result of a deterministic simulation of Equation (1). a) At a low photon rate of $\lambda_0 T_d$ = 12.5, the dynamics appear Poissonian. The time series has 
 no visible structure, the autocorrelation function is sharply peaked at 0, and the distribution of photon counts in a window of $w=T_d/4$ is nearly Poisson. b) $\lambda_0 T_d$ = 200,
 a slow modulation of the photon rate is evident. c) at  $\lambda_0 T_d$ = 3200, the photon rate varies smoothly, the photon count distribution is bimodal and much wider than a Poisson distribution with the 
 same mean, and the autocorrelation function shows slow oscillations. The deterministic simulation d) shows the same features as the high 
 photon rate data shown in c).}\label{fig2}      
\end{figure}
Figure 2 shows several time series recorded with this system with increasing photon rate, showing    
a transition from Poisson noise to deterministic chaos.
We plot $N_w(t)$, the number of photon arrivals in the interval [$t-w$,$t$]. In Figure 2, all of the 
plots were generated with $w=T_d/4$. When the incident photon 
rate is $\lambda_0 T_d = 12.5$, the photons appear to arrive at random, uncorrelated times as in a stationary Poisson process.
Increasing the incident photon rate to $\lambda_0 T_d = 200$, a smooth modulation of the photon rate starts to become apparent. 
At $\lambda_0 T_d = 3200$, $N_w(t)$ has a smooth character, and qualitatively resembles a low-dimensional chaotic signal. 
We also plot the results of a deterministic simulation using equation (\ref{detEOM}). This time series was smoothed 
with a moving average over a time window of width $w$ to be directly comparable with $N_w(t)$. 
We plot the autocorrelation function, $C(t')= \left \langle (N_w(t)-\bar{N_w})(N_w(t-t')-\bar{N_w}) \right \rangle$,
normalized so that the value of the autocorrelation function is unity at $t'=0$. 
As the photon rate increases from $\lambda_0 T_d = 12.5$, the autocorrelation function changes from a $\delta$-like peak, 
characteristic of a Poisson process, to an oscillatory function which shows correlations at long timescales (tens of milliseconds). The autocorrelation function 
of the deterministic simulation time series is in close agreement with the autocorrelation function of the photon arrivals with $\lambda_0 T_d = 3200$.
Histograms of $N_w(t)$ also show a transition from a nearly Poisson distribution to a bimodal distribution characteristic of
the deterministic chaotic process. 
\begin{figure}
 %\centering
 \includegraphics{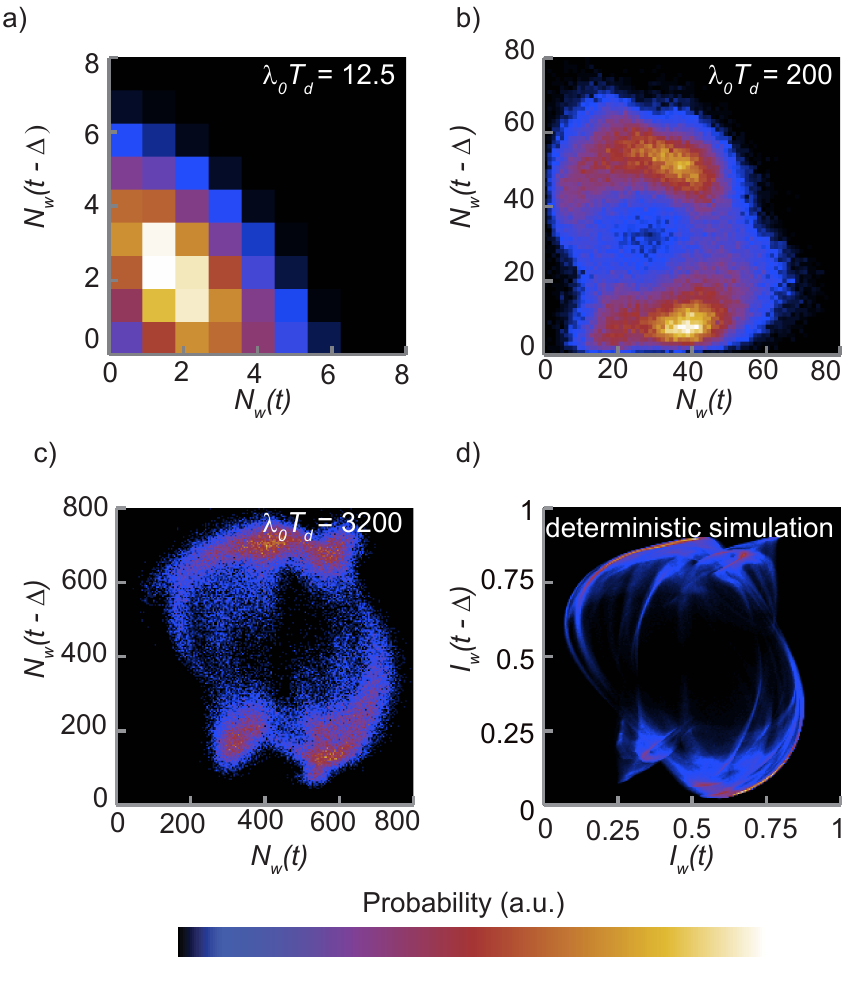}
 \caption{Poincar\'{e} sections. We visualize the emergence of a chaotic attractor from 
 Poisson noise with increasing photon rate by embedding photon count time series in two dimensions with a time delay of
 $\Delta=T_d/4$, and reducing the dimensionality of the dynamics by plotting points only when the state of the system in phase space passes though a codimension-1 surface defined by
 $x_1-x_2=\pi$. a) - c) show experimental data, and d) shows the result of a deterministic simulation. These histograms are constructed with a bin width of 1 photon in in a) and b), 4 photons in c), and 0.005 in d)}\label{fig3}
\end{figure}

To visualize the development of chaos with increasing photon rate, we show Poincar\'{e} surfaces of section in Figure 3.
We perform a time-delay embedding of the experimental time series $N_w(t)$, using a time delay of
$\Delta=T_d/4$, by constructing a list of points in two dimensional space of the form $\left[ N_w(t), N_w(t-\Delta) \right ]$.
Because the attractor has a dimension higher than 2, we reduce the dimensionality of the attractor by plotting the points only when the state variables
pass through a codimension 1 Poincar\'{e} surface defined by $x_1-x_2=\pi$. The embeddings show a similar trend to the plots 
in Figure 2. We see a development of complex chaotic dynamics from discrete photon noise as the photon rate increases. The 
deterministic simulation is plotted for comparison, and, as in Figure 2, a moving average of width $w$ is employed so that
the smoothed intensity time series, $I_w(t)$, is directly analogous to $N_w(t)$. The deterministic signal in
Figure 3d can be regarded as the infinite photon rate limit of the photon counting system. 

Figure 4 shows the dependence of the variance of $N_w$ on the window $w$ and offers another indication of the transition from shot noise to deterministic chaos. 
The time integral of an uncorrelated random signal executes a random walk in which the variance grows linearly with the integration time. For this reason,
we plot $\Var(N_w)/w$ in Figure 4. We see distinct asymptotic growth rates of the variance with small and large $w$. When $w$ is small, the variance 
reflects the Poissonian nature of the photon arrivals, and the growth rate of the variance has roughly constant value of $\Var(N_w)/w=\lambda_0\bar{I}$. 
In the limit where the counting window is much longer than the time scale of the variations in intensity, $N_w(t)$ can be regarded as the sum of the photon counts in many 
independent identically distributed intervals, and the central limit theorem implies that the variance will grow in proportion to $w$.
As we increase the photon rate from $\lambda_0 T_d = 12.5$ to  $\lambda_0 T_d = 3200$, we see an increasing offset between the two asymptotic rates of growth of the variance. 
The variance can be related to the photon rate, counting window, and the unnormalized autocorrelation function, 
$c_I(t')=\left \langle (I(t)-\bar{I})(I(t-t')-\bar{I}) \right \rangle$ \cite{mandel1995}.
\begin{eqnarray} 
\Var(N_w)=w \left[  \lambda_0\bar{I} + \lambda^2_0 \underbrace{\int_{-w}^{w} dt' \left( 1-\frac{|t'|}{w}\right)c_I(t')}_{\Theta(w)} \right] \label{vareq}
\end{eqnarray}
The second term in Equation (\ref{vareq}) accounts for the difference between the observed variance and the variance of a Poisson process with the same rate.
The quantity $\Theta(w)$ has units of time, and measures the correlations in $I(t)$ introduced by the feedback. This quantity increases from 0 to an asymptotic 
value $\Theta_\infty$ as $w$ increases, accounting for the shape of the curves shown in Figure 4. In deterministic simulations, we find $\Theta_\infty$ = 150 $\mu$s.
The value of $\Theta_\infty$ is related to the size of the intensity fluctuations, and the rate at which $\Theta(w)$ approaches this asymptotic value is determined by the timescales
of the correlations of $I(t)$. 

\begin{figure}%[htp]
 %\centering
 \includegraphics{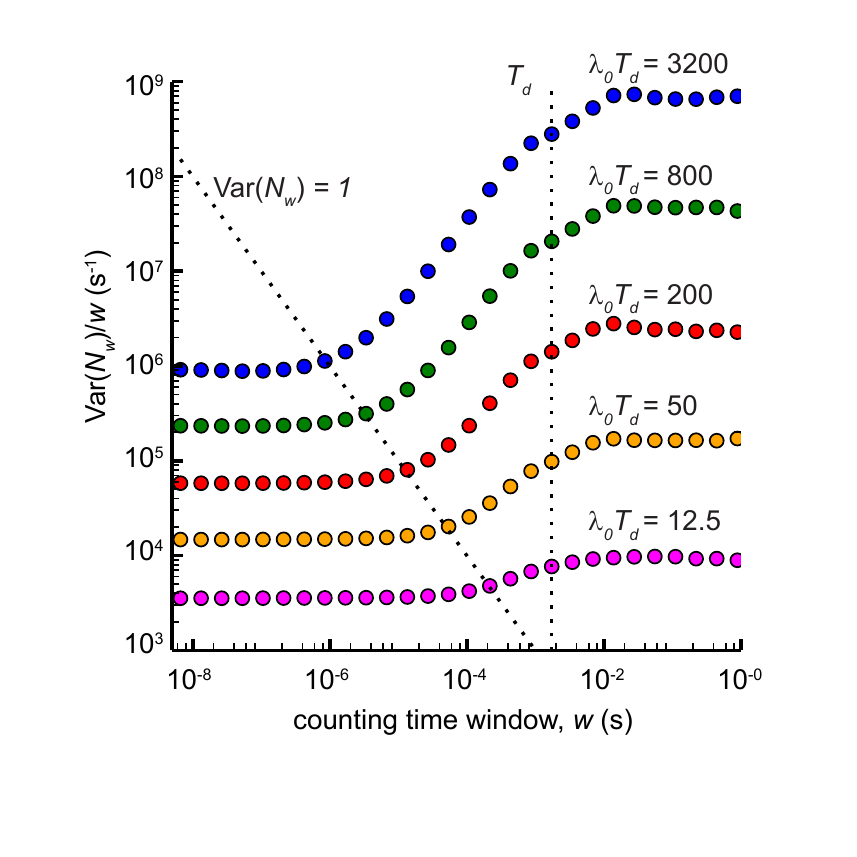}
 \caption{Experimental dependence of growth rate of variance on counting time window, $w$. To indicate the timescale of the deterministic dynamics we indicate $T_d$. 
 We also show a line indicating $\Var(N_w)=1$, which roughly separates timescales over which less than one photon arrives in the counting window from timescales over which many photons arrive. 
 Distinct asymptotic values of $\Var(N_w)/w$ are seen in the limits of small and large $w$. The offset between these two values reflects deterministic correlations
 in the photon arrival rate and grows with increasing photon rate.}\label{fig4}
\end{figure}

\subsection{Entropy quantification}

We characterize the entropy production using the $(\varepsilon,\tau)$ entropy per unit time, $h(\varepsilon, \tau)$ \cite{gaspard1993}. This 
measurement of entropy has two parameters: sample resolution,  $\varepsilon$, and
sampling time interval $\tau$, which are natural parameters for most experiments because measurement devices record data to finite resolution at discrete times. In addition to being 
experimentally relevant, the dependence of $h(\varepsilon,\tau)$ on these parameters can reflect the underlying physical origin of 
unpredictability\cite{gaspard1993, cencini2000, boffetta2002}. 

In chaotic systems, unpredictability is due to the sensitive dependence on initial conditions. Because small perturbations grow exponentially in time, chaotic systems generate information. 
The growth of uncertainty is quantified by the Lyapunov exponents $\mu_i$, and in particular the largest exponent, $\mu_1$. Positive Lyapunov exponents and entropy rate both 
quantify unpredictability and there is a close relationship between these two quantities. One would expect that if a chaotic system is sampled infrequently ($\tau \mu_1 \gg 1$), 
successive samples will be uncorrelated because of the growth of uncertainty between measurements. On the other hand, if the interval between successive samples is small ($\tau \mu_1 \ll 1)$, 
one expects strong correlations between adjacent samples and a reduced entropy per sample. Experimental and theoretical work
using semiconductor lasers has shown that these considerations are crucial to physical random number
generation using chaotic dynamics \cite{harayama2012,sunada2012,mikami2012}. In the limit that $\tau, \varepsilon \rightarrow 0$, $h(\varepsilon,\tau)$ will approach a finite value,
the Kolmogorov-Sinai (or metric) entropy, $h_{ks}$ \cite{pesin1977,eckmann1985,benettin1976,boffetta2002}. The metric entropy is related to the Lyapunov exponents, $\mu_i$, by
\begin{equation}
h_{ks}=\frac{1}{\log(2)} \sum_{\mu_i>0} \mu_i. \label{pesin}
\end{equation}
We calculated the spectrum of Lyapunov exponents from Equation (1) \cite{farmer1982, geist1990}. There is only one positive Lyapunov exponent 
with a value of $\mu_1/\log(2)$ = 345 bits/s. The Kaplan-Yorke dimension\cite{grassberger1983} calculated from the Lyapunov spectrum is 3.56. 

The $(\varepsilon,\tau)$ entropy will have qualitatively different behavior as $\varepsilon \rightarrow 0$ depending on the physical origin of unpredictability.
In chaotic systems, the entropy rate does not depend on either the sampling rate or the sampling resolution. This property of chaotic systems
imposes a theoretical limitation on physical random number generation. Increasing the speed and resolution of a measurement device
cannot in principle increase the entropy that can be harvested from a deterministic chaotic system beyond $h_{ks}$. In contrast to deterministic systems, 
the, entropy rate of stochastic signals diverges like  $-\log(\varepsilon)$ for finite $\tau$\cite{gaspard1993, cencini2000}.

Another advantage of the $(\varepsilon,\tau)$ entropy is that it can be calculated from experimental data using an algorithm described by Cohen and Procaccia\cite{cohen1985}.
In our case, we chose to calculate the entropy from $N_w(t)$ with a counting time window of $w=T_d/4$. With this window, $N_w(t)$
approximates the behavior of the deterministic signal $I(t)$ as seen in Figures 2 and 3. We do not employ an averaging time window to 
compute the entropy from deterministic simulations.

The first step to computing the entropy rate of an experimental signal is to generate a list of points in $d$-dimensional space using time delay embedding 
with a delay of $\tau$. These vectors can be regarded as samples of a $d$-dimensional probability distribution over phase space. 
The entropy of this probability distribution, $H_d$, is sometimes referred to as the pattern entropy for patterns of length $d$ \cite{gaspard1998}.
In principle $H_d$ can be calculated by building a histogram with boxes of width $\varepsilon$ and applying Shannon's formula, $H=-\sum_i p_i \log_2 p_i$ \cite{cover2012}. 
In practice, direct application of this approach requires a very large amount of data when the embedding dimension is large. Cohen and Procaccia proposed a 
more efficient algorithm to estimate the pattern entropy \cite{cohen1985} in the context of estimating metric entropy from experimental data. 
First, one randomly selects a small number $M$ of reference points from the time 
series. In our case, $M=5000$ was sufficient. For each reference point $i$, one computes $n_i (\varepsilon)$, the fraction of points within 
a box of width $\varepsilon$ centered on the reference point. The only difference between a direct calculation of the Shannon entropy, and 
the Cohen-Procaccia procedure is that in a direct calculation, a rectangular array of bins is used, rather than a set of bins centered on 
random points chosen from the data set. In searching for neighbors for the $i$th reference point, we exclude 
points within a time window of $\tau$ of that point, as suggested by Theiler \cite{theiler1990}.
The pattern entropy is then estimated by 
\begin{equation}
H_d(\varepsilon) = -\frac{1}{M}  \sum_{i=1}^{M} \log_2  n_i(\varepsilon). \label{Hd}
\end{equation}
It is a general feature of unpredictable signals that $H_d$ grows linearly with $d$ in the limit that $d$ is large, and the entropy rate is the slope of this linear increase. 
\begin{equation}
h(\varepsilon,\tau) = \frac{1}{\tau} \lim\limits_{d\rightarrow\infty} \left[ H_{d}(\varepsilon) - H_{d-1}(\varepsilon) \right] \label{hdiff}
\end{equation}

\begin{figure}%[htp]
 \centering
 \includegraphics{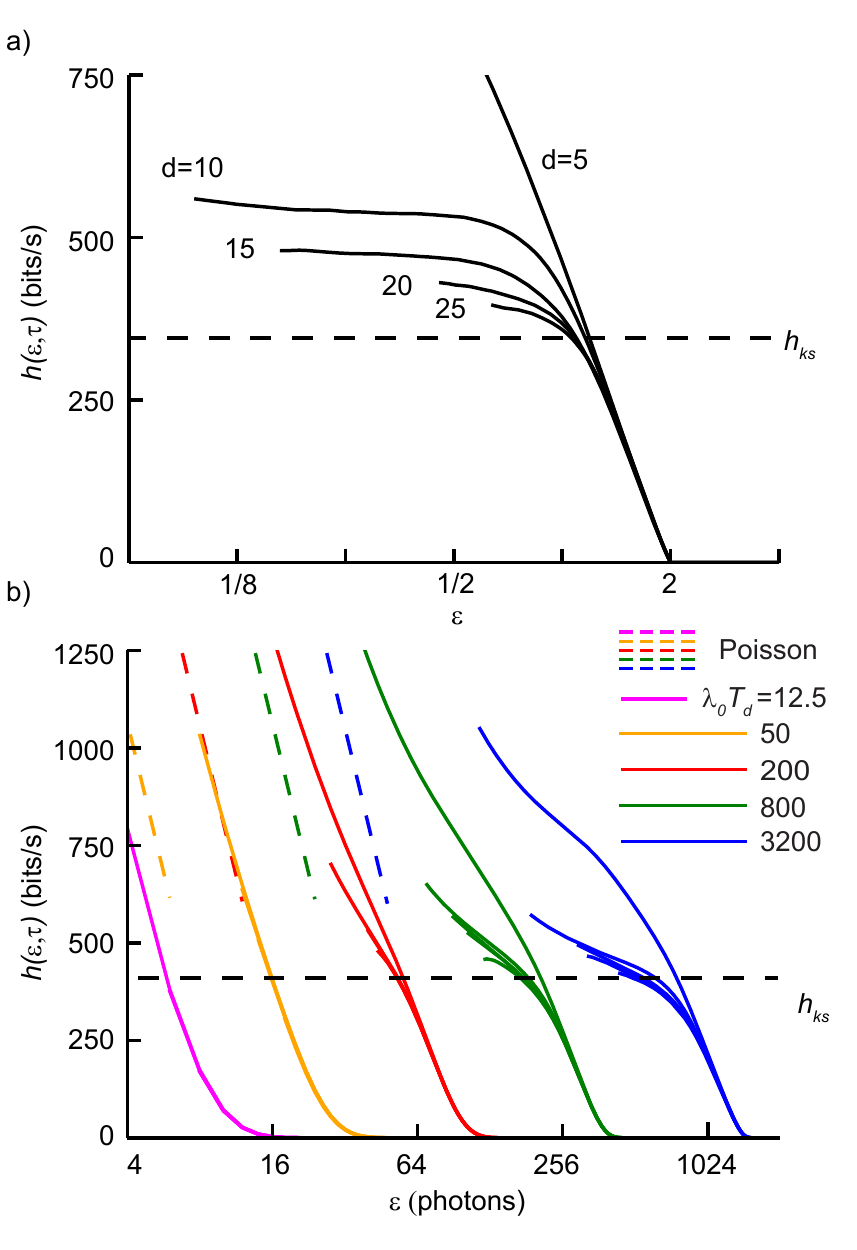}
 \caption{$(\varepsilon, \tau)$ entropy in deterministic simulation and experiment. Calculations with embedding 
 dimensions of 5,10,15,20 and 25 are shown. a) Deterministic simulation. The entropy is calculated from the intensity, $I(t)$, 
 which is normalized so that its values are between 1(complete transmission through the modulator) and 0. $\varepsilon$ 
 is measured in the same units as $I(t)$. Characteristic of deterministic systems, the entropy rate is independent of $\varepsilon$, 
 and approaches the largest Lyapunov exponent as $d$ increases. b) Entropy rate of experimental time series. 
 The entropy is calculated from $N_w(t)$, and $\varepsilon$ is measured in photons. At low photon rates we see a divergence 
 characteristic of a Poisson processes. As the photon rate increases, the dependence of the entropy on $\varepsilon$ becomes progressively 
 flatter, and approaches $h_{ks}$. Across photon rates, we see a divergence of the entropy rate for small $\varepsilon$. 
 The Poisson curves were calculated by approximating the Shannon entropy of a Poisson process by an integral over a Gaussian distribution with a mean and variance of $\lambda_0 \bar{I} w$. 
 }\label{fig5}
\end{figure}

Figure 5 shows the entropy per unit time in both deterministic simulation and experiment with $\tau=(3/4) T_d$. The duration of the simulation was $512 \times 10^6$ $T_d$. Figure 5a 
shows that in the deterministic simulation, the entropy rate remains flat as $\varepsilon$ decreases. As $d$ increases, 
this plateau approaches $h_{ks}$, as indicated by Equation (\ref{pesin}). In Figure 5b, we see that as the photon rate increases, the dependence of the entropy on $\varepsilon$
becomes progressively flatter at high $\varepsilon$. Furthermore, in the region that this flattening is present, the value of $h(\varepsilon,\tau)$ 
is close to $h_{ks}$. The flattening of $h(\varepsilon,\tau)$ at high photon rate is another indication that this system 
behaves more deterministically in this regime. At all photon rates, we see $h(\varepsilon,\tau)$ sharply increases as $\epsilon$ decreases,
which is due to the shot noise inherent in the system. It is natural to compare the entropy rates we observe to a constant-rate Poisson process 
with the same average rate. The Poisson curves in Figure 5 were calculated by approximating the Shannon entropy of 
a Poisson distribution by an integral over a Gaussian distribution with a mean and variance of $\lambda_0 \bar{I} w$.
In the limit that $\varepsilon \ll \sqrt{\lambda_0\bar{I} w}$, this leads to the asymptotic expression 
$h(\varepsilon,\tau) = (-1/\tau) \log_2 \left(\varepsilon/\sqrt{2\pi e \lambda_0 \bar{I} w} \right)$, 
indicated by the dashed curves in Fig. 5(b). For small $\varepsilon$, $h(\varepsilon,\tau)$ increases 
logarithmically with decreasing $\varepsilon$, and parallels this curve. 
This logarithmic dependence is more pronounced at lower $\lambda_0$.  

\section{Discussion}

We show in this paper that the choice of the resolution with which we observe our system
allows us to see either noisy or deterministic dynamics. By counting photon arrivals over timescales 
on the order of the delay time and filter time constants, we see deterministic dynamics in
the time series, Poincar\'{e} sections, and the autocorrelation functions. 
Furthermore, when we observe the dynamics on large scales of both value ($\varepsilon$) and time ($w$ and $\tau$), we find that the
entropy rate is close in value to the metric entropy calculated from the positive Lyapunov exponents of the deterministic model, 
which shows that the entropy generation is dominated by the deterministic exponential amplification of small perturbations in this regime.

In contrast, by employing high resolution in photon counts and time scales, we see that 
both the entropy rate and variance reflect the stochastic nature of the photon arrivals. For small values of $w$, the variance of the number 
of photon counts is equal to the average number of counts, characteristic of a Poisson process.
The logarithmic dependence of the entropy on $\varepsilon$ shown in Figure 5 offers another indication of the noisy nature of the dynamics
at small scales. In addition to showing both shot noise and chaos at different scales, our experiment also shows a transition from shot noise to chaos with increasing photon rate. 
The precise control over the rate of photon arrivals and dynamical timescales afforded by our experiment allows for experimental observation of the interplay of noise and dynamics. Our results can 
be seen to bridge two widely used methods of physical random number generation.  

Two prevalent methods have attracted attention for optical random number generation:  those based on single photon 
detection from strongly attenuated light sources \cite{jennewein2000, Dynes2008}, and those based on digitized high-speed fluctuations
from chaotic lasers \cite{uchida2008}.   In the former case, the entropy is claimed to originate entirely from quantum mechanical uncertainty,
yet in practice these methods are also subject to unpredictable drift and environmental variations. 
In the latter case, the entropy is attributed to the dynamical unpredictability of chaos, but the unavoidable presence of spontaneous emission
is thought to play a role in seeding these macroscopic fluctuations \cite{harayama2012, sunada2012}.  The system presented here is unprecedented 
in that it can approach macroscopic chaos from the single photon limit, thereby revealing the transition from noise to chaos.  
Moreover, the analysis offers a unified measure of entropy that captures both behaviors, and clarifies the relationship between sampling 
frequency, measurement resolution and entropy rate.

The designer of a physical random number generator must choose the sampling rate and resolution that they will use to collect numbers
from a physical system. These decisions will impact the entropy rate. Heuristically, finer discretization (smaller $\varepsilon$) and more 
frequent sampling (smaller $\tau$) lead to higher entropy rates, but without the methods presented here it is difficult to assess the dependence of 
the entropy rate on these parameters in any given system. The statistical tests that are usually used to evaluate physical random number 
generation \cite{rukhinnist2001,marsaglia1998diehard} were not designed to answer these questions, but rather to certify that a stream of 
bits is free of bias and correlation. If a random number generator employs post processing (as most do), existing statistical tests applied
to the output binary sequence provide no insight into whether the entropy originates from the physical process or the post-processing algorithm employed. 
The $(\varepsilon,\tau)$ entropy clarifies the origin and nature of uncertainty and informs the choice of sampling rate and measurement resolution.

\begin{acknowledgments}
We gratefully acknowledge grant N000141410443 from the Office of Naval Research. 
The authors acknowledge the University of Maryland supercomputing resources (http://www.it.umd.edu/hpcc) made available in conducting the research reported in this paper. 
\end{acknowledgments}

\bibliography{bibliography}

\end{article}

\end{document}